\documentclass[11pt,onecolumn,amssymb,nofootinbib]{revtex4} 
\usepackage{amsmath}
\usepackage{amssymb}
\usepackage{babel}
\begin{document}

\title{Collapse dynamics are diffusive}

\author{Sandro Donadi}
\email{s.donadi@qub.ac.uk}
\affiliation{Centre for Quantum Materials and Technologies, School of Mathematics and Physics, Queen’s University, Belfast BT7 1NN, United Kingdom}

\author{Luca Ferialdi}
\email{luca.ferialdi@gmail.com}
\affiliation{Department of Physics and Chemistry, University of Palermo, via Archirafi 36, I-90123 Palermo, Italy}

\author{Angelo Bassi}
\email{abassi@units.it}
\affiliation{Department of Physics, University of Trieste, Strada Costiera 11, 34151 Trieste, Italy
 \\ INFN,
Sezione di Trieste, Strada Costiera 11, 34126 Trieste, Italy}

\begin{abstract}
Non-interferometric experiments have been successfully employed to constrain models of spontaneous wave function collapse, which predict a violation of the quantum superposition principle for large systems. These experiments are grounded on the fact that, according to these models, the dynamics is driven by a noise that, besides  collapsing the wave function in space,  generates a diffusive motion with characteristic signatures, which, though small, can be tested. The non-interferometric approach might seem applicable only to those models which implement the collapse through a noisy dynamics, not to any model, which collapses the wave function in space. Here we show that this is not the case: under reasonable assumptions, {\it any collapse dynamics (in space) is diffusive}. Specifically, we prove that any space-translation covariant dynamics which complies with the no-signaling constraint, if collapsing the wave function in space, must change  the average momentum of the system, and/or its  spread. 
\end{abstract}

\maketitle

{\it Introduction.}
Searching for the potential limits of validity of the quantum superposition principle is of highest relevance  for the foundations of Quantum Mechanics and in general for our understanding of Nature \cite{leggett1980macroscopic,Leggett871,bell2004speakable,weinberg2012collapse}, and more pragmatically also for the scalability of quantum technologies. One further motivation is  provided by models of spontaneous wave function collapse \cite{ghirardi1986unified,ghirardi1990continuous,bassi2003dynamical,bassi2013models}, which  predict a progressive breakdown of quantum linearity and the localization in space of physical systems as their size increases, thus justifying the emergence of a classical world from quantum constituents.

The most natural way of testing the superposition principle is via {\it interferometric experiments}, and several  platforms have been employed or proposed for this scope, including  atoms \cite{kasevich1991atomic, kovachy2015quantum}, molecules \cite{arndt1999wave,fein2019quantum}, optomechanical systems \cite{bateman2014near}, crystals \cite{belli2016entangling}, nitrogen-vacancy centers \cite{scala2013matter}. The difficulty in performing such experiments is that it is hard to generate and maintain a macroscopic spatial superposition in a (almost) decoherence-free environment, and check whether over time it survives or decays. The current world record for a matter wave delocalized in space is about 0.5 m, obtained with cold atoms \cite{kovachy2015quantum}, while the largest mass which interfered with itself weights about $10^5$ amu \cite{fein2019quantum}. 
We are still far away from probing the quantum nature of the macroscopic world, but impressive technological development makes the goal less far away in the future \cite{pino2018chip,gasbarri2021testing,belenchia2021test}.

Meanwhile, a different strategy has been successfully employed to test models of spontaneous wave function collapse, which consists of {\it non-interferometric experiments} \cite{carlesso2022present}. The basic principle lies in the observation that in all these models the collapse of the wave function is triggered by a noise, which shakes the particles' dynamics, whether their wave function is localized or not. As such, particles undergo a characteristic diffusion  process, which can be tested through high-precision position measurements; these, although demanding, are easier to perform than the interferometric ones. Examples of these kind of experiments are the precise monitoring of the motion of cold atoms \cite{bilardello2016bounds}, cantilevers \cite{vinante2016upper, vinante2020narrowing} or gravitational wave detectors \cite{carlesso2016experimental,helou2017lisa}. Another consequence of the collapse-induced diffusion is that atoms emit an extra radiation, which is not predicted by standard quantum mechanics; also this effect has been  used to test collapse models \cite{fu1997spontaneous,adler2007photon,adler2013spontaneous,donadi2014spontaneous}. The bounds on the phenomenological parameters of the Continuous Spontaneous Localization (CSL) model \cite{ghirardi1990continuous} resulting from non-interferometric experiments are six orders of magnitude stronger than the best bounds set by interferometric experiments~\footnote{The best bound from interferometric experiments comes from macromolecule interferometry and for the reference value $r_c=10^{-7}$ m it is $\lambda\leq 10^{-7}$ s$^{-1}$ (see fig. 3b of \cite{fein2019quantum}). This should be compared to the bound coming from X-rays emission from Germanium which is $\lambda\leq 5.2 \times 10^{-13}$ s$^{-1}$ (see fig. 4 of \cite{donadi2021novel})}. Moreover, a recent experiment based on precise measurements of the radiation emitted from Germanium ruled out the parameters free version of the  Di\'osi-Penrose (DP) model \cite{donadi2020underground}.

An apparently weak side of non-interferometric experiments is that they seem not to represent a direct test of the quantum superposition principle, but only of the models so far proposed, which explicitly violate it. This leaves  the possibility open to formulate a model where the wave function does collapse, without inducing diffusion on the system. Here we show that this is not possible: any dynamics that localizes the wave function in space also changes the momentum.  We call this {\it diffusion}, because in all current collapse models it manifests as such, but in this paper it is meant to signify any change in the momentum of the system. Since a change in momentum can be (at least in principle) detected by non-interferometric experiments,  our result shows that they represent a test of the quantum superposition principle in a stronger sense than one might suppose.

We consider a general situation:  we assume that physical systems are associated with a wave function $\psi$, which is subject to a generic norm-preserving  (possibly, non linear) dynamics. The requirements on the dynamics are: {\it i}) it does not allow for superluminal signaling; {\it ii}) it is space-translation covariant, at least at the statistical level. The first assumption implies that the dynamics for the wave function $\psi$ also provides a well-defined dynamics for the density matrix $\hat\rho$, which in general is not true for a nonlinear evolution \cite{gisin1989stochastic}.  Then, by construction, the dynamics for $\hat\rho$ is linear, completely positive and trace preserving (see Supplementary Material below). The second assumption amounts to requiring that the map for $\hat\rho$ is space-translation covariant and has the same physical motivation on which all physical fundamental theories are based. We will prove our result for a single particle, where with particle we also (and mostly) mean the center of mass of a composite system.

To fix the notation, we will consider a particle in a box of size $L$, with periodic boundary conditions; this choice will avoid potential problems when dealing with plane waves. Let $\hat{p}_i$ be the momentum operator along direction $i\; (= x,y,z)$, and $(2\pi\hbar/L) n_i$  its eigenvalues, with $n_i \in {\mathbb{Z}}$; let $\boldsymbol{n} = (n_x, n_y, n_z)$. The average value of the momentum operator for a given state $\hat\rho$ is denoted as $\overline{p}_{i,\hat\rho} = \text{Tr}(\hat{p}_i \hat\rho) $ and its variance as $\Delta p_{i,\hat\rho}= \text{Tr}(\hat{p}^2_i\hat \rho) - [\text{Tr}( \hat{p}_i\hat \rho)]^2$. Last, let $\hat{\boldsymbol{n}} = |\boldsymbol{n}\rangle\langle\boldsymbol{n}|$ be the state of definite momentum $(2\pi\hbar/L) \boldsymbol{n}$.

We will prove the following {\it theorem}: Consider a particle in a box of size $L$ with periodic boundary conditions for its wave function. Consider a  dynamical map for the wave function satisfying the conditions {\it i}) and {\it ii}), and let $\varPhi$ be the associated map for the density matrix $\hat\rho$ \footnote{We assume $\hat{\rho}$ belonging to the domain of $\hat{p}_i$ and $\hat{p}^2_i$, and such that its image under $\varPhi$ belongs to the same domain.}. Assume that the average momentum is conserved along the three directions: $\overline{p}_{i,\varPhi[\hat\rho]} = \overline{p}_{i,\hat\rho}$ for any $\hat\rho$. Then $\Delta p_{i,\varPhi[\hat{\boldsymbol{n}}]} = \Delta p_{i,\hat{\boldsymbol{n}}}$ $\forall \boldsymbol{n}$ if an only if $\varPhi$ is a function  of the momentum operator only, in which case plane waves do not collapse in space. Moreover, if $\Delta p_{i,\varPhi[\hat{\boldsymbol{n}}]} \neq 0$ for some $\boldsymbol{n}$, then $\Delta p_{i,\varPhi[\hat{\rho}]} > \Delta p_{i,\hat{\rho}}$ for any $\hat\rho$ such that  $\langle \boldsymbol{n}| \hat\rho | \boldsymbol{n} \rangle \neq 0$. 

Before proceeding with the proof, some comments are at order. We assume that the average momentum $\overline{p}_{i}$ is conserved for all states because, if this is not true for a certain $ \hat\rho$, then a non-interferometric experiment is immediately available, namely to measure the change of $\overline{p}_{i}$ induced by $\varPhi$ on the $\hat\rho$. Most collapse models in the literature conserve the average momentum. An exception are the dissipative models, as for example the dissipative CSL \cite{smirne2015dissipative}; however, also in this case, for generic states $\Delta p_{i,\hat\rho}$ changes due to the collapse i.e. there is diffusion. This is discussed in detail in the Supplementary Material C below.

Plane waves are the most delocalized states, and a sensible collapse model is expected to collapse them in space. In that case, $\Delta p_i$ changes for all of them (it is 0 before the collapse, and not 0 after); then the theorem tells that $\Delta p_i$ will  increase for any $\hat\rho$. This means that any sensible collapse model must induce an  increase of $\Delta p_i$ of the system, for any state (delocalized or not). When the map acts repeatedly over time, the increase of $\Delta p_i$ amounts to some form of diffusion. 

Our proof is valid for free particles as well as for particles interacting with an external potential. In the second case, in general, the change in $\Delta p_i$ cannot be easily separated into the contribution coming from the interaction potential and that coming from the collapse. However, all experiments are such that this separation is possible, either because the particle is free or because the effect of the interaction potential can be estimated.

The theorem might seem a manifestation of Heisenberg's uncertainty principle: a collapse in position must increase the spread in momentum; this is true for plane waves, but it is not necessary when the state is not a minimum uncertainty state (the vast majority of them are not). Yet the theorem says that also in that case the spread in momentum  increases~\footnote{See also the final comment in section C of the Supplementary Material below}. 

{\it Proof.}
As discussed in the supplementary material below, the map $\varPhi$ is linear, trace preserving and completely positive. Then Kraus' theorem \cite{kraus1983states} states that it is of the form:
\begin{equation} 
    \varPhi[\hat{\rho}]=\sum_{k}\hat{A}_{k}\hat{\rho} \hat{A}_{k}^{\dagger}\label{kraus}
\end{equation} 
where  the operators $ \hat{A}_k$  satisfy the condition $\sum_k \hat{A}_{k}^{\dagger}\hat{A}_{k}=1$. 

 The structure of translation covariant maps is characterized by Holevo's theorem~\cite{holevo1998radon}, whose essence is the following.
A map is covariant under a space translation amounting to a displacement $\boldsymbol{x}$ if, for any $\hat{\rho}$: 
\begin{equation}
e^{-\frac{i}{\hbar}\hat{\boldsymbol{p}}\cdot\boldsymbol{x}}\,\varPhi[\hat{\rho}]\,e^{\frac{i}{\hbar}\hat{\boldsymbol{p}}\cdot\boldsymbol{x}}=\varPhi\left[e^{-\frac{i}{\hbar}\hat{\boldsymbol{p}}\cdot\boldsymbol{x}}\,\hat{\rho}\, e^{\frac{i}{\hbar}\hat{\boldsymbol{p}}\cdot\boldsymbol{x}}\right]\,;\label{c1}
\end{equation}
by multiplying the two sides of this equation on the left by $e^{\frac{i}{\hbar}\hat{\boldsymbol{p}}\cdot\boldsymbol{x}}$ and on the right by $e^{-\frac{i}{\hbar}\hat{\boldsymbol{p}}\cdot\boldsymbol{x}}$ and using Eq.~\eqref{kraus}, one finds that:
\begin{equation}
\varPhi[\hat{\rho}]=\sum_{k}\hat{A}_{k}(\boldsymbol{x})\hat{\rho} \hat{A}_{k}^{\dagger}(\boldsymbol{x})\label{inv1}
\end{equation}
with 
\begin{equation}\label{Ax}
\hat{A}_{k}(\boldsymbol{x})=e^{\frac{i}{\hbar}\hat{\boldsymbol{p}}\cdot\boldsymbol{x}}\hat{A}_{k}e^{-\frac{i}{\hbar}\hat{\boldsymbol{p}}\cdot\boldsymbol{x}}.
\end{equation}

By requiring the covariance in Eq. (\ref{c1}) to hold for any possible displacement $\boldsymbol{x}$ with $x_j\in[-L/2,L/2]$, one eventually finds:
\begin{equation}
\varPhi[\hat{\rho}]=\frac{1}{L^{3}}\int_{-\frac{L}{2}}^{+\frac{L}{2}}d\boldsymbol{x}\sum_{k}A_{k}(\boldsymbol{x})\hat{\rho} A_{k}^{\dagger}(\boldsymbol{x})\label{Kraus in box}
\end{equation}
with $A_{k}(\boldsymbol{x})$ given by Eq. (\ref{Ax}) and
\begin{equation}
\sum_{\boldsymbol{m}}\sum_{k}|\langle\boldsymbol{m}|A_{k}|\boldsymbol{n}\rangle|^{2}=1\qquad\boldsymbol{m},\boldsymbol{n}\in\mathbb{Z}^{3}\,.\label{this}
\end{equation}
Eq.~\eqref{Kraus in box} represents the general space translation covariant Kraus map inside a box.

Since we are assuming that the average momentum does not change, the change of its spread is given by 
\begin{equation}\label{Dmaintext}
    D_{i,\hat{\rho}}:=\text{Tr}(\hat{p}_{i}^{2}\varPhi[\hat{\rho}])-\text{Tr}(\hat{p}_{i}^{2}\hat{\rho}),
\end{equation}
which, according to Eq. (\ref{Kraus in box}), is equal to (see Supplementary material below):
\begin{equation}
D_{i,\hat{\rho}}=\sum_{\boldsymbol{m},\boldsymbol{n}}P(\boldsymbol{m},\boldsymbol{n})\tilde{m}_{i}^{2}\langle\boldsymbol{n}|\hat{\rho}|\boldsymbol{n}\rangle\label{Yj},
\end{equation}
where $\tilde{m}_{i}:=(2\pi\hbar/L)m_i$ and
\begin{equation}\label{P}
    P(\boldsymbol{m},\boldsymbol{n}):=\sum_{k}|\langle\boldsymbol{m}+\boldsymbol{n}|A_{k}|\boldsymbol{n}\rangle|^{2}.
\end{equation}
The requirement that the map $\Phi$ does not lead to diffusion is equivalent to asking that  $D_{i, \hat{\rho}} = 0$ for any $\hat{\rho}$. We now prove that a map fulfilling this condition cannot collapse the wave function in space.
By assuming that $D_{i,\hat{\rho}}=0$ for any statistical operator of the form $\hat{\rho}=|\boldsymbol{n}_{0}\rangle\langle\boldsymbol{n}_{0}|$, we conclude that:
\begin{equation}
\sum_{\boldsymbol{m}}P(\boldsymbol{m},\boldsymbol{n}_{0})\tilde{m}_{i}^{2}=0\qquad\forall\boldsymbol{n}_{0}.\label{cond2new}
\end{equation}

At this point, it is convenient to introduce the marginal distributions of $P(\boldsymbol{m},\boldsymbol{n}_{0})$ given by:
\begin{equation}
    P_i(m_i,\boldsymbol{n}_{0}):=\sum_{m_{j\neq i}} P(\boldsymbol{m},\boldsymbol{n}_{0})\,\label{marg},
\end{equation}
which allow to rewrite Eq. (\ref{cond2new}) as:
\begin{align}
\sum_{m_i}P_i(m_i,\boldsymbol{n}_{0})m_{i}^{2}=0.\label{cond2new2}
\end{align}
for all $\boldsymbol{n}_{0}$.  Eqs.~\eqref{this} and \eqref{P} imply that $P(\boldsymbol{m},\boldsymbol{n})$ is a probability distribution for the variables $\boldsymbol{m}$, for any fixed $\boldsymbol{n}$. It follows from Eq.~\eqref{cond2new2} that the marginals $P_i(m_i,\boldsymbol{n}_{0})$ are probability distributions with zero variance, which implies:
\begin{equation}
P_{i}(m_{i},\boldsymbol{n}_{0})=\delta_{m_{i},0}\qquad \forall \boldsymbol{n}_{0}.
\end{equation}
Since this is true for all the marginals of $P(\boldsymbol{m},\boldsymbol{n}_0)$, then:
\begin{equation}\label{finalP}
P(\boldsymbol{m},\boldsymbol{n})=\sum_{k}|\langle\boldsymbol{m}+\boldsymbol{n}|A_{k}|\boldsymbol{n}\rangle|^{2}=\delta_{\boldsymbol{m},0}\qquad \forall \boldsymbol{n}.
\end{equation}
The equation above implies that:
\begin{equation}\label{ccc}
|\langle\boldsymbol{m}+\boldsymbol{n}|A_{k}|\boldsymbol{n}\rangle|^{2}=c_{k}(\boldsymbol{n})\delta_{\boldsymbol{m},0}\qquad\forall\boldsymbol{n} \,,
\end{equation}
with  $c_{k}(\boldsymbol{n})$ generic non-negative functions such that $\sum_{k}c_{k}(\boldsymbol{n})=1$. 
By writing the matrix element in the form $\langle\boldsymbol{m}+\boldsymbol{n}|\hat A_{k}|\boldsymbol{n}\rangle=R_{k}(\boldsymbol{m},\boldsymbol{n})e^{i\varphi_{k}(\boldsymbol{m},\boldsymbol{n})}$ one finds that
$ R_{k}(\boldsymbol{m},\boldsymbol{n})=\sqrt{c_{k}(\boldsymbol{n})}\delta_{\boldsymbol{m},0}$
and therefore
\begin{eqnarray}\label{Akfinal}
\hat A_{k}=\!\sum_{\boldsymbol{m},\boldsymbol{n}}|\boldsymbol{m}+\boldsymbol{n}\rangle\langle\boldsymbol{m}+\boldsymbol{n}|\hat A_{k}|\boldsymbol{n}\rangle\langle\boldsymbol{n}|=\!\sum_{\boldsymbol{n}}\!\sqrt{c_{k}(\boldsymbol{n})}e^{i\varphi_{k}(0,\boldsymbol{n})} |\boldsymbol{n}\rangle\langle\boldsymbol{n}|\!= \!\hat{A}_k(\hat {\boldsymbol p}), \;\;\;\;
\end{eqnarray}
where the last equality signifies that the operators $\hat A_{k}$ are functions of the momentum operator only.
As such, the map $\varPhi$ becomes:
\begin{equation}\label{final_kraus_nocoll}
    \varPhi[\hat \rho] = \sum_k \hat A_k(\hat {\boldsymbol p}) \hat{\rho} \hat A_k^{\dagger}(\hat {\boldsymbol p}).
\end{equation}

Typical examples of maps of this kind contain only one Kraus operator and are the free evolution ($\hat A( \hat{\boldsymbol p}) \sim \exp[-i\hat p^2/2m]$) and spatial translations ($\hat A(\hat{\boldsymbol p}) \sim \exp[i \hat{\boldsymbol p} \cdot {\boldsymbol a}]$), which do not modify the spread in momentum. 

It is trivial to check that a map like~\eqref{final_kraus_nocoll} does not change the momentum distribution, and  that plane waves $\hat{\rho}=|\boldsymbol{n}_{0}\rangle\langle\boldsymbol{n}_{0}|$ are stationary states,
 which implies that the map $\varPhi$ is not capable of collapsing such fully delocalized states in position. 

Coming to the second part of the theorem, let us assume that the map changes the spread in momentum  of a given  momentum eigenstate $|\boldsymbol{n_0}\rangle$; this implies that:
\begin{equation}
    \sum_{\boldsymbol{m}}P(\boldsymbol{m},\boldsymbol{n}_{0})\tilde{m}_{i}^{2}>0.
\end{equation}
Then from Eq. (\ref{Yj}) it is clear that for all states $\hat{\rho}$ such that $\langle \boldsymbol{n_0}|\hat{\rho}|\boldsymbol{n}_{0}\rangle \neq 0$  the spread in momentum also increases under the action of the map. This completes the proof.

The proof presented here is applied to a single quantum operation, but clearly holds for a sequence of them. A proof of the theorem for Lindblad's dynamics is given in the Supplementary Material below. 


{\it Discussion.}
The no-faster-than-light-signaling requirement, which implies that the dynamical map for the density matrix is uniquely identified and is linear, sets a very strong constraint on the possible collapse dynamics. In particular, it tells that the dynamics must be such that, at the {\it statistical level} (i.e. $\varPhi$), it acts on the density matrix in the same way whether the underlying mixture is made of delocalized states (which should collapse) or of localized states (which are not expected to further collapse). This excludes the possibility of having a collapse dynamics that takes place only when the system is in a superposition, and is suspended when the system is not; in this second case the effect might be small or null for some specific states, but the dynamics as such is there. 

Since the time when a superposition is created is not specified {\it a priori}, and since $\varPhi$ is ``blind'' to the states forming a statistical mixture, the dynamics must act repeatedly in time with a sufficiently high rate, to make sure that superpositions do not live too long (in the Markovian limit, one typically has a Lindblad dynamics).

Note that such dynamics implies a time directionality (see \cite{albert2001time} for further discussion). At the statistical level this is clear since pure states evolve into statistical mixtures. At the wave function level, time directionality arises because spatial superpositions collapse to localised states, while the opposite does not occur. The collapse occuring in position implies energy non-conservation.

Space translation covariance enters as follows. Consider (in one dimension) a partition of the real line into small enough intervals and let $\hat{A}_k$ be the  projection operators associated to the intervals. The associated Kraus map $\varPhi$ in Eq.~\eqref{kraus}  is not covariant under space translations. When applied to a generic superposition, it collapses it and changes the spread in momentum, while preserving its average; when the map is applied any other time, the state does not change anymore (here we are neglecting the Hamiltonian dynamics). In this case diffusion does not occur, apart for the change in momentum in the very first instance. Space translation covariance requires that there are no privileged points in space, so no privileged partition of space; therefore no state, however localized (except for the pathological---and unstable under the free evolution---case of a Dirac delta), can remain unaltered by a repeated application of the map, because there is always the chance that it is further localized by the action of an operator $\hat{A}_k$ associated to an interval which does not entirely contain the state when the map acts. This is the source of diffusion.

We implicitly framed our theorem in a non-relativistic setting, but we do not see any fundamental obstacle in extending it to a relativistic scenario.

{\it Conclusions.} Modifying the quantum dynamics provided by the Schr\"odinger equation is tricky and easily generates nonphysical situations. One of the most common problems is superluminal  signaling with arbitrarily high speed; collapse models are designed to avoid this problem. Our theorem shows why the no-signaling constraint, together with space translation covariance,  requires that collapse in position always comes with diffusion. For this reason, non-interferometric experiments are equally good as interferometric ones for testing these models; as anticipated in the introduction, the first type of tests are easier to perform and have already set significantly stronger bounds on the collapse parameters, ruling out some of them. 

The same logic applies to any open-quantum-system dynamics: typical environments induce decoherence in position~\footnote{See~\cite{adler2003decoherence} for the difference between collapse models and decoherence in resolving the quantum measurement problem.} and the resulting dynamics is space translational covariant, because most interactions depend on the relative distances among particles---all fundamental ones do; as such they must generate also diffusion. This is reflected by the Lindblad structure of the most common master equations: when the Lindblad operators depend on position, the expectation value of $\hat p^2$ is not constant. 

This fact has consequences, for example, regarding recent proposals for searching for dark matter signals using matter-wave interferometers \cite{riedel2013direct, bateman2015existence, alonso2022cold}, which are sensitive to the decoherence induced by dark matter particles. Equally well, one can propose non-interferometric experiments~\cite{bateman2015existence}, which are sensitive to the diffusion generated by  the particles; the application of non-interferometric techniques to collapse models proved that they hold the potential of providing stronger results. 

{\it Acknowledgements.} The authors thank S. L. Adler, L. Diósi, A. Smirne, H. Ulbricht, and B. Vacchini for several useful comments. S. D. acknowledges the financial support from the Marie Sklodowska Curie Action through the UK Horizon Europe guarantee administered by UKRI and from INFN. L. F. acknowledges support from MUR through project PRIN (Project No. 2017SRN-BRK QUSHIP). A. B. acknowledges financial support from the University of Trieste, INFN, the EIC Pathfinder project QuCoM (GA No. 101046973), and the PNRR MUR Project No. PE0000023-NQSTI.

\bibliography{biblio.bib}{}
\bibliographystyle{apsrev4-1}

\newpage

\begin{widetext}

\section{Supplementary Material}
\subsection{Maps for the wave function and for the density matrix}\label{ap1}

Let us consider a ($N$ dimensional, for simplicity) Hilbert space $\mathbb H$ associated to the physical system of interest, and a generic norm-preserving map $T: \mathbb H \rightarrow \mathbb H$, which might not be linear. The map might be stochastic, as typical of collapse models; we will denote the stochastic average with a bar over the quantities to be averaged over the noise (if the map is deterministic, the bar can be removed from the proof). 

Let us also consider the associated space $\mathcal{B}(\mathbb H)$ of $N \times N$ matrices, to which the density matrix belongs. A density matrix $\hat{\rho}$ in general admits multiple decompositions in terms of pure states, for example:
\begin{equation}
    \hat{\rho} = \sum_n p_n |\psi_n\rangle\langle \psi_n| = \sum_k q_k |\phi_k \rangle\langle \phi_k |,\label{SM}
\end{equation}
where $\{ |\psi_n\rangle \}$ and $\{ |\phi_k\rangle \}$ are two different sets of normalized vectors in $\mathbb H$, while $\{ p_n \}$ and $\{ q_k \}$ are two sets of positive real numbers summing to 1, which represent our ignorance about the precise state of the system. The two statistical mixtures $\{ (|\psi_n\rangle, p_n) \}$ and $\{ (|\phi_k\rangle, q_k) \}$ are said to be {\it equivalent} since they correspond to the same statistical operator as per Eq.~\eqref{SM}. 

A generic map $T$ need not preserve the equivalence between statistical mixtures, i.e.
\begin{equation}
    \hat{\rho}'_{(1)} \equiv \sum_n p_n \overline{ |T\psi_n\rangle\langle T\psi_n|} \neq \hat{\rho}'_{(2)} \equiv \sum_k q_k \overline{ |T\phi_k \rangle\langle T\phi_k |},
\end{equation}
where $|T\psi_n\rangle = T[|\psi_n\rangle]$ and similarly for $|\phi_k \rangle$. However, if this happens it can be shown that, under the further assumption that measurements occur (either effectively or fundamentally) as provided by the Born rule and the von Neumann projection postulate, one can establish a protocol for superluminal signaling \cite{gisin1989stochastic,bassi2015no}. Rejecting this possibility amounts to asking that the map $T$ preserves the equivalence among statistical mixtures. In such a case, the map induces a map among density matrices according to:
\begin{eqnarray} \label{eq:rtyhrst}
    \Lambda : \mathcal{B}(\mathbb H) & \rightarrow & \mathcal{B}(\mathbb H) \nonumber \\
    \hat{\rho} & \rightarrow & \Lambda [\hat{\rho}] = \sum_n p_n \overline{ |T\psi_n\rangle\langle T\psi_n|},
\end{eqnarray}
where $\{ (|\psi_n\rangle, p_n) \}$ is now {\it any} statistical mixture associated to $\hat{\rho}$.

We now show that the map $\Lambda$ thus defined is {\it linear}, {\it  positive} and {\it trace preserving}. Linearity is proven as follows: suppose that $\hat{\rho}$ is the convex sum of $\hat{\rho}_1$ and $\hat{\rho}_2$ according to: $\hat{\rho} = p_1 \hat{\rho}_1 + p_2 \hat{\rho}_2$, with $p_1, p_2 \geq 0$ and $p_1 + p_2 = 1$; consider also a statistical mixture $\{ (|\psi_{1n}\rangle, p_{1n}) \}$ associated to $\hat{\rho}_1$ and $\{ (|\psi_{2n}\rangle, p_{2n}) \}$ associated to $\hat{\rho}_2$. Then:
\begin{eqnarray}
\Lambda [\hat{\rho}] & = & \Lambda\left[\sum_{j=1}^{2}p_{j}\sum_{n}p_{jn}|\psi_{jn}\rangle\langle\psi_{jn}|\right]= \nonumber \\
& = &  \sum_{j=1}^{2}p_{j}\sum_{n}p_{jn} \overline{ |T\psi_{jn}\rangle\langle T\psi_{jn}|} = p_1 \sum_n p_{1n} \overline{ |T\psi_{1n}\rangle\langle T\psi_{1n}|} + p_2 \sum_n p_{2n} \overline{ |T\psi_{2n}\rangle\langle T\psi_{2n}|} \nonumber \\
& = & p_1 \Lambda \left[ \sum_n p_{1n} |\psi_{1n}\rangle\langle \psi_{1n}| \right] + p_2 \Lambda \left[ \sum_n p_{2n} |\psi_{2n}\rangle\langle \psi_{2n}| \right] \nonumber \\
& = &  p_1 \Lambda[\hat{\rho}_1] + p_2 \Lambda[\hat{\rho}_2];
\end{eqnarray}
in the first line we simply rewrote $\hat{\rho}$ in terms of the statistical mixture  $\{( |\psi_{1n}\rangle, p_{1n} )\} \cup \{ (|\psi_{2n}\rangle, p_{2n}) \}$; in going from the first to the second line we used Eq.~\eqref{eq:rtyhrst} applied to $\hat{\rho}$; in going from the second to the third line we used again Eq.~\eqref{eq:rtyhrst}, this time applied to $\hat{\rho}_1$ and $\hat{\rho}_2$; in the last line, we used the fact that $\{ (|\psi_{1n}\rangle, p_{1n}) \}$ and $\{ (|\psi_{2n}\rangle, p_{2n}) \}$ are two statistical mixtures associated to $\hat{\rho}_1$ and $\hat{\rho}_2$ respectively.

The map $\Lambda$ is automatically positive since it maps wave functions into wave functions; it is also trace preserving, given that it maps statistical mixtures into statistical mixtures. Now we discuss  complete positivity.

 A map $\Lambda$ on $\mathcal{B}(\mathbb H)$ is completely positive if, for any $M \in \mathbb N$, the extended map $I_{\text{\tiny M}} \otimes \Lambda$ on $\mathcal{B}(\mathbb H_{\text{\tiny M}} \otimes\mathbb H)$ is positive, where $I_{\text{\tiny M}}$ is the identity map on $\mathcal{B}(\mathbb H_{\text{\tiny M}})$ and $\mathbb H_{\text{\tiny M}}$ is a $M$-dimensional Hilbert space. The Hilbert space $\mathbb H_{\text{\tiny M}}$  refers to any additional degree of freedom, which is not affected by the considered dynamics.
Asking for an ancilla to exist, which is not affected by the considered dynamics, is in principle an additional assumption; anyhow, in all collapse models so far formulated, such an ancilla naturally exists in a strong sense (for example the spin of fermions is not affected by the collapse) and in a weak sense (systems like photons can have an arbitrarily weak coupling to the collapse noise, if their energy is arbitrarily low).
 
 The map $\tilde{T}: \mathbb H_{\text{\tiny M}} \otimes\mathbb H  \rightarrow \mathbb H_{\text{\tiny M}} \otimes\mathbb H$ must exist, otherwise by simply  attaching the system of interest to an ancilla, there would be no dynamics for the wave function anymore; it must also preserve the equivalence among statistical mixtures, for the same reason spelled above. Then $\tilde{T}$ generates the map $\tilde{\Lambda}: \mathcal{B}(\mathbb H_{\text{\tiny M}} \otimes\mathbb H) \rightarrow \mathcal{B}(\mathbb H_{\text{\tiny M}} \otimes\mathbb H)$ according to:
\begin{eqnarray} \label{eq:rtyhrssrdy}
    \tilde\Lambda : \mathcal{B}(\mathbb H_{\text{\tiny M}} \otimes\mathbb H) & \rightarrow & \mathcal{B}(\mathbb H_{\text{\tiny M}} \otimes\mathbb H) \nonumber \\
    \tilde{\hat{\rho}} & \rightarrow & \tilde\Lambda\; [ \tilde{\hat{\rho}}] =  \sum_n \tilde{p}_n \overline{|\tilde T \tilde\psi_n\rangle\langle \tilde T \tilde\psi_n|},
\end{eqnarray} 
with obvious meaning of symbols; again, $ \tilde\Lambda$ is linear.

Since $T$ on $\mathbb {H}$ is in general nonlinear, its extension $\tilde T$ on $\mathbb H\otimes\mathbb H_{\text{\tiny M}}$ (which  is also nonlinear) is in general not uniquely determined by $T$, even if the ancilla does not evolve. However, locality, which is an instance of non-faster-than-light signaling, requires that, for {\it factorized} states,  the map $\tilde{T}$ factorizes into:
\begin{eqnarray}
    I_{\text{\tiny M}} \times T : \mathbb H_{\text{\tiny M}} \times\mathbb H  & \rightarrow &  \mathbb H_{\text{\tiny M}} \times\mathbb H \nonumber \\
    |\phi\rangle \otimes  |\psi\rangle & \rightarrow &  |\phi\rangle \otimes  |T\psi\rangle
\end{eqnarray}
(the ancilla could be arbitrarily far away). This map generates the following map among {\it factorized} density matrices:
\begin{eqnarray} \label{eq:rtyhrstgdfd}
    I_{\text{\tiny M}} \times \Lambda : \mathcal{B}(\mathbb H_{\text{\tiny M}}) \times \mathcal{B}(\mathbb H) & \rightarrow & \mathcal{B}(\mathbb H_{\text{\tiny M}}) \times \mathcal{B}(\mathbb H) \nonumber \\
    \rho_{\text{\tiny M}} \otimes \hat{\rho} & \rightarrow & (I_{\text{\tiny M}} \times \Lambda)\;  [\hat{\rho}_{\text{\tiny M}} \otimes \hat{\rho}] = \hat{\rho}_{\text{\tiny M}} \otimes \Lambda[\hat \rho],
\end{eqnarray}
which can be extended by linearity to the whole tensor product space, thus defining:
\begin{eqnarray} \label{eq:rtyhrstgdfdd}
    I_{\text{\tiny M}} \otimes \Lambda : \mathcal{B}(\mathbb H_{\text{\tiny M}} \otimes \mathbb H) & \rightarrow & \mathcal{B}(\mathbb H_{\text{\tiny M}} \otimes \mathbb H).
\end{eqnarray}
The two maps $\tilde \Lambda$ and $I_{\text{\tiny M}} \otimes \Lambda$ are the same map, since they are both linear and coincide on a basis of $\mathcal{B}(\mathbb H_{\text{\tiny M}} \otimes \mathbb H)$. Since $\tilde \Lambda$ is positive by construction, because it maps statistical mixtures in statistical mixtures, also $I_{\text{\tiny M}} \otimes \Lambda$ is. This proves that $\Lambda$ is completely positive. 

 As a final note, we point out that in \cite{caiaffa2017stochastic} it was shown that also positive but non completely-positive dynamics admit stochastic unravelings. 
 This means that the stochastic unraveling $T$, which is not associated to a completely-positive dynamics, does not admit an extension $\tilde T$ to a larger Hilbert space, like the one discussed here above.

\subsection{Derivation of Eq. (8) in the main text}\label{app_der1011}
In this section we derive the formulas for the expectation values of $\hat{p}_{j}$ and $\hat{p}_{j}^2$ with respect to $\varPhi[\hat{\rho}]$. The reason why we include also the average momentum is that it is relevant for the analysis carried on in appendix \ref{app_gen}. We start from Eq. (5) in the main text, and we compute 
\[
\textrm{Tr}\left\{\hat{p}_{j}\varPhi[\hat{\rho}]\right\}=\frac{1}{L^{3}}\int_{-\frac{L}{2}}^{+\frac{L}{2}}d\boldsymbol{x}\sum_{k}\sum_{\boldsymbol{n}}\langle\boldsymbol{n}|\hat{p}_{j}\hat{A}_{k}(\boldsymbol{x})\hat{\rho}\hat{A}_{k}^{\dagger}(\boldsymbol{x})|\boldsymbol{n}\rangle=
\]
\begin{equation}
=\frac{1}{L^{3}}\int_{-\frac{L}{2}}^{+\frac{L}{2}}d\boldsymbol{x}\sum_{k}\sum_{\boldsymbol{n},\boldsymbol{\ell},\boldsymbol{m}}\langle\boldsymbol{n}|\hat{p}_{j}e^{\frac{i}{\hbar}\hat{\boldsymbol{p}}\cdot\boldsymbol{x}}A_{k}e^{-\frac{i}{\hbar}\hat{\boldsymbol{p}}\cdot\boldsymbol{x}}|\boldsymbol{\ell}\rangle\langle\boldsymbol{\ell}|\hat{\rho}|\boldsymbol{m}\rangle\langle\boldsymbol{m}|e^{\frac{i}{\hbar}\hat{\boldsymbol{p}}\cdot\boldsymbol{x}}A_{k}^{\dagger}e^{-\frac{i}{\hbar}\hat{\boldsymbol{p}}\cdot\boldsymbol{x}}|\boldsymbol{n}\rangle.
\end{equation}
By exploiting $
\hat{\boldsymbol{p}}|\boldsymbol{n}\rangle=(2\pi\hbar/L)\boldsymbol{n}|\boldsymbol{n}\rangle$ we find
\begin{align}
\textrm{Tr}\left\{\hat{p}_{j}\varPhi[\hat{\rho}]\right\}&=\frac{2\pi\hbar}{L}\sum_{k}\sum_{\boldsymbol{n},\boldsymbol{\ell},\boldsymbol{m}}n_{j}\underset{\langle\boldsymbol{\ell}|\boldsymbol{m}\rangle}{\underbrace{\left(\frac{1}{L^{3}}\int_{-\frac{L}{2}}^{+\frac{L}{2}}d\boldsymbol{x}e^{\frac{2\pi i}{L}(\boldsymbol{m}-\boldsymbol{\ell})\cdot\boldsymbol{x}}\right)}}\langle\boldsymbol{n}|A_{k}|\boldsymbol{\ell}\rangle\langle\boldsymbol{\ell}|\hat{\rho}|\boldsymbol{m}\rangle\langle\boldsymbol{m}|A_{k}^{\dagger}|\boldsymbol{n}\rangle=\nonumber
\\
&=\frac{2\pi\hbar}{L}\sum_{\boldsymbol{n},\boldsymbol{\ell}}n_{j}\sum_{k}|\langle\boldsymbol{n}|A_{k}|\boldsymbol{\ell}\rangle|^{2}\langle\boldsymbol{\ell}|\hat{\rho}|\boldsymbol{\ell}\rangle=\frac{2\pi\hbar}{L}\sum_{\boldsymbol{n},\boldsymbol{\ell}}\widetilde{P}(\boldsymbol{n},\boldsymbol{\ell})n_{j}\langle\boldsymbol{\ell}|\hat{\rho}|\boldsymbol{\ell}\rangle\label{bo}
\end{align}
where we defined 
\begin{equation}
\widetilde{P}(\boldsymbol{n},\boldsymbol{\ell}):=\sum_{k}|\langle\boldsymbol{n}|A_{k}|\boldsymbol{\ell}\rangle|^{2}.
\end{equation}
Performing the change of variables $\boldsymbol{m}=\boldsymbol{n}-\boldsymbol{\ell}$ and then relabelling $\boldsymbol{\ell}\rightarrow\boldsymbol{n}$ we finally get 
\begin{equation}
\textrm{Tr}\left\{\hat{p}_{j}\varPhi[\hat{\rho}]\right\}=\textrm{Tr}\left\{\hat{p}_{j}\hat{\rho}\right\}+\frac{2\pi\hbar}{L}\sum_{\boldsymbol{m},\boldsymbol{n}}P(\boldsymbol{m},\boldsymbol{n})m_{j}\langle\boldsymbol{n}|\hat{\rho}|\boldsymbol{n}\rangle
\end{equation}
where 
\begin{equation}
P(\boldsymbol{m},\boldsymbol{n})=\widetilde{P}(\boldsymbol{m}+\boldsymbol{n},\boldsymbol{n})=\sum_{k}|\langle\boldsymbol{m}+\boldsymbol{n}|A_{k}|\boldsymbol{n}\rangle|^{2}.
\end{equation}
is precisely Eq. (9) of the main text. 

Recalling that $\tilde{m}_j=\frac{2\pi\hbar}{L}m_{j}$, it follows that:
\begin{equation}\label{Xj}
d_{j,\hat{\rho}}:=\textrm{Tr}\left\{\hat{p}_{j}\varPhi[\hat{\rho}]\right\}-\textrm{Tr}\left\{\hat{p}_{j}\hat{\rho}\right\}=\sum_{\boldsymbol{m},\boldsymbol{n}}P(\boldsymbol{m},\boldsymbol{n})\tilde{m}_{j}\langle\boldsymbol{n}|\hat{\rho}|\boldsymbol{n}\rangle\,,
\end{equation}

A similar calculation leads to 
\begin{equation}
\textrm{Tr}\left\{\hat{p}_{j}^{2}\,\varPhi[\hat{\rho}]\right\}=\frac{(2\pi\hbar)^{2}}{L^{2}}\sum_{\boldsymbol{n},\boldsymbol{\ell}}\widetilde{P}(\boldsymbol{n},\boldsymbol{\ell})n_{j}^{2}\langle\boldsymbol{\ell}|\hat{\rho}|\boldsymbol{\ell}\rangle
\end{equation}
Then again by performing the change of variables $\boldsymbol{m}=\boldsymbol{n}-\boldsymbol{\ell}$ and then relabelling $\boldsymbol{\ell}\rightarrow\boldsymbol{n}$ we get:
\begin{align}\label{last}
\textrm{Tr}\left\{\hat{p}_{j}^{2}\,\varPhi[\hat{\rho}]\right\}&=\frac{(2\pi\hbar)^{2}}{L^{2}}\sum_{\boldsymbol{m},\boldsymbol{n}}P(\boldsymbol{m},\boldsymbol{n})\left(m_{j}+n_{j}\right)^{2}\langle\boldsymbol{n}|\hat{\rho}|\boldsymbol{n}\rangle
\\
&=\textrm{Tr}\left\{\hat{p}_{j}^{2}\,\hat{\rho}\right\}+\sum_{\boldsymbol{m},\boldsymbol{n}}P(\boldsymbol{m},\boldsymbol{n})\tilde{m}_{j}^{2}\langle\boldsymbol{n}|\hat{\rho}|\boldsymbol{n}\rangle+2\sum_{\boldsymbol{m},\boldsymbol{n}}P(\boldsymbol{m},\boldsymbol{n})\tilde{m}_{j}\langle\boldsymbol{n}|\hat{p}_{j}\hat{\rho}|\boldsymbol{n}\rangle,\nonumber
\end{align}
which implies
\begin{equation}\label{Dappendix}
    D_{j,\hat{\rho}}=\textrm{Tr}\left\{\hat{p}_{j}^{2}\,\varPhi[\hat{\rho}]\right\}-\textrm{Tr}\left\{\hat{p}_{j}^{2}\,\hat{\rho}\right\}=\sum_{\boldsymbol{m},\boldsymbol{n}}P(\boldsymbol{m},\boldsymbol{n})\tilde{m}_{j}^{2}\langle\boldsymbol{n}|\hat{\rho}|\boldsymbol{n}\rangle+2\sum_{\boldsymbol{m},\boldsymbol{n}}P(\boldsymbol{m},\boldsymbol{n})\tilde{m}_{j}\langle\boldsymbol{n}|\hat{p}_{j}\hat{\rho}|\boldsymbol{n}\rangle.
\end{equation}
When $d_{j,\hat{\rho}}=0$ for all $\hat{\rho}$, by taking $\hat{\rho}=|\boldsymbol{n}_0\rangle\langle\boldsymbol{n}_0|$ one finds $\sum_{\boldsymbol{m}}P(\boldsymbol{m},\boldsymbol{n}_{0})\tilde{m}_{j}=0$ for any $\boldsymbol{n}_0$; In such a case the last term of Eq. (\ref{Dappendix}) is always equal to zero, from which Eq. (8) in the main text follows. 
\subsection{General proof of the theorem, with $d_{j,\hat{\rho}}\neq 0$ }\label{app_gen}

In the main text, we proved the theorem with the simplifying assumption $d_{j,\hat{\rho}}=0$. Here we want to generalise the first part of the theorem also to maps that change the average momentum. Typical examples of this kind of maps are dissipative dynamics; in the context of collapse models, dissipative extensions of the Ghirardi-Rimini-Weber (GRW) model, CSL model and QMUPL model were introduced in \cite{bassi2005energy,ferialdi2012dissipativePRA,ferialdi2012dissipativePRL,smirne2014dissipative,smirne2015dissipative}. Non-interferometric tests of the dissipative CSL model were studied in \cite{nobakht2018unitary}.  

We will prove the following theorem: consider a map of the form in Eq. (5) in the main text (i.e. fulfilling conditions ({\it i}) and ({\it ii}) of the main text); consider the difference in the momentum spread after and before the application of the map i.e. 
\begin{equation}\label{diff_spread}
    \Delta_{j,\hat{\rho}}:=\Delta p_{i,\varPhi[\hat{\rho}]} - \Delta p_{i,\hat{\rho}}=D_{j,\hat{\rho}}-d_{j,\hat{\rho}}^{2}-2\langle\hat{p}_j\rangle d_{j,\hat{\rho}}
\end{equation}
where $\Delta p_{j,\hat\rho}= \text{Tr}(\hat{p}^2_j\hat \rho) - [\text{Tr}( \hat{p}_j\hat \rho)]^2$ and $d_{j,\hat{\rho}}$ and $D_{j,\hat{\rho}}$ are defined, respectively, in Eqs. (\ref{Xj}) and (\ref{Dappendix}); If any state of the form: 
\begin{equation}\label{rho}
\hat{\rho}=|\psi\rangle\langle\psi|\;\;\;\;\textrm{with}\;\;\;\;|\psi\rangle=a|\boldsymbol{n}_{0}\rangle+b|\boldsymbol{m}_{0}\rangle
\end{equation}
(where $a,b$ are generic complex coefficients which satisfy the normalization condition $|a|^{2}+|b|^{2}=1$) satisfy $\Delta_{j,\hat{\rho}}=0$ (no spread in the momentum) then the map $\varPhi$ is such that: 
\begin{equation}
    \varPhi[|\boldsymbol{n}_{0}\rangle\langle\boldsymbol{n}_{0}|]=|\boldsymbol{\gamma}(\boldsymbol{n}_{0})+\boldsymbol{n}_{0}\rangle\langle\boldsymbol{\gamma}(\boldsymbol{n}_{0})+\boldsymbol{n}_{0}|
\end{equation}
with $\boldsymbol{\gamma}(\boldsymbol{n}_{0})\in\mathbb{Z}^{3}$. This implies that this map does not collapse the momentum eigenstates (it just acts on them as a boost) hence it cannot be a satisfactory dynamics for the wave function collapse in space.

{\it Proof.} It is convenient to introduce the following notation:
\begin{equation}\label{mrn}
\overline{m_j(\boldsymbol{n})}:=\sum_{\boldsymbol{m}}P(\boldsymbol{m},\boldsymbol{n})\tilde{m}_j\qquad \overline{m^2_j(\boldsymbol{n})}:=\sum_{\boldsymbol{m}}P(\boldsymbol{m},\boldsymbol{n})\tilde{m}^2_j\,,
\end{equation}
which exploits the fact that, for each $\boldsymbol{n}$, $P(\boldsymbol{m},\boldsymbol{n})$
is a probability distribution of the variable $\boldsymbol{m}$,
in such a way that $\overline{m_j(\boldsymbol{n})}$ and  $\overline{m^{2}_j(\boldsymbol{n})}$ represent averages with respect
to it. 

In this new notation Eqs. (\ref{Xj}) and (\ref{Dappendix}) become:
\begin{equation}
d_{j,\hat{\rho}}=\sum_{\boldsymbol{n}}\overline{m_j(\boldsymbol{n})}\langle\boldsymbol{n}|\hat{\rho}|\boldsymbol{n}\rangle,
\end{equation}
\begin{equation}
D_{j,\hat{\rho}}=\sum_{\boldsymbol{n}}\left(\overline{m^{2}_j(\boldsymbol{n})}+2\overline{m_{j}(\boldsymbol{n})}\tilde{n}_j\right)\langle\boldsymbol{n}|\hat{\rho}|\boldsymbol{n}\rangle.\label{Y-2}
\end{equation}
Given a state of the form in Eq. (\ref{rho}) one has:
\begin{equation}
\langle\boldsymbol{n}|\hat{\rho}|\boldsymbol{n}\rangle=|a|^{2}\delta_{\boldsymbol{n},\boldsymbol{n}_{0}}+|b|^{2}\delta_{\boldsymbol{n},\boldsymbol{m}_{0}}
\end{equation}
By choosing $a=1$ and $b=0$ one gets
\begin{equation}
\Delta_{j,\hat{\rho}}=\overline{m_j^{2}(\boldsymbol{n}_{0})}-\left(\overline{m_j(\boldsymbol{n}_{0})}\right)^{2}
\end{equation}
and requiring $\Delta_{j,\hat{\rho}}=0$ implies: 
\begin{equation}
\overline{m_j^{2}(\boldsymbol{n}_{0})}=\left(\overline{m_j(\boldsymbol{n}_{0})}\right)^{2}.\label{cond1-2}
\end{equation}
We can now compute $\Delta_{j,\hat{\rho}}$ for generic coefficients $a$ and $b$ using
condition (\ref{cond1-2}). This leads to:
\begin{equation}
\Delta_{j,\hat{\rho}}=|a|^{2}|b|^{2}\left(\overline{m_j(\boldsymbol{n}_{0})}-\overline{m_j(\boldsymbol{m}_{0})}\right)\left[\left(\overline{m_j(\boldsymbol{n}_{0})}-\overline{m_j(\boldsymbol{m}_{0})}\right)-2\left(\tilde{m}_{0j}-\tilde{n}_{0j}\right)\right]
\end{equation}
and the condition $\Delta_{j,\hat{\rho}}=0$ implies:
\begin{equation}
\overline{m_j(\boldsymbol{n}_{0})}-\overline{m_j(\boldsymbol{m}_{0})}=\begin{cases}
0\\
2\left(\tilde{m}_{0j}-\tilde{n}_{0j}\right)
\end{cases}.\label{COND2-1}
\end{equation}

In is now convenient to rewrite Eqs.~\eqref{mrn}
as follows:
\begin{equation}
\overline{m_{j}(\boldsymbol{n}_{0})}=\sum_{m_{j}}P_{j}(m_{j},\boldsymbol{n}_{0})\tilde{m}_{j}
\qquad
\overline{m_{j}^{2}(\boldsymbol{n}_{0})}=\sum_{m_{j}}P_{j}(m_{j},\boldsymbol{n}_{0})\tilde{m}_{j}^{2}
\end{equation}
where $P_j$ are the marginals defined in Eq.~(11) of the main text. As a consequence, condition (\ref{cond1-2}) reads
\begin{equation}
\sum_{m_{j}}P_{j}(m_{j},\boldsymbol{n}_{0})m_{j}^{2}-\left(\sum_{m_{j}}P_{j}(m_{j},\boldsymbol{n}_{0})m_{j}\right)^{2}=0\,,
\end{equation}
which implies that each $P_{j}(m_{j},\boldsymbol{n}_{0})$ is a distribution with
zero variance, i.e. 
\begin{equation}
P_{j}(m_{j},\boldsymbol{n}_{0})=\delta_{m_{j},\gamma_{j}(\boldsymbol{n}_{0})}\label{P_delta}
\end{equation}
By replacing this identity in Eq. (\ref{COND2-1}) one finds
\begin{equation}
\overline{m_{j}(\boldsymbol{n}_{0})}-\overline{m_{j}(\boldsymbol{m}_{0})}=\frac{2\pi\hbar}{L}\left(\gamma_{j}(\boldsymbol{n}_{0})-\gamma_{j}(\boldsymbol{m}_{0})\right)=\begin{cases}
0\\
-2\frac{2\pi\hbar}{L}\left(n_{0j}-m_{0j}\right)
\end{cases}
\end{equation}
for all $\boldsymbol{n}_{0},\boldsymbol{m}_{0}$. This implies that
either: 
\begin{equation}
\gamma_{j}(\boldsymbol{n}_{0})=\gamma_{j}(\boldsymbol{m}_{0})\;\Longrightarrow\;\gamma_{j}(\boldsymbol{n}_{0})=\gamma_{j}\label{gamma1}
\end{equation}
or 
\begin{equation}
\gamma_{j}(\boldsymbol{n}_{0})-\gamma_{j}(\boldsymbol{m}_{0})=-2\left(n_{0j}-m_{0j}\right).\label{gamma2}
\end{equation}
Since this equation must hold for all $\boldsymbol{n}_{0}$
and $\boldsymbol{m}_{0}$, it follows that $\gamma_{j}(\boldsymbol{n}_{0})$ must
depend only on the component $n_{0j}$, which in turn implies 
\begin{equation}
\gamma_{j}(\boldsymbol{n}_{0})=\gamma_{j}-2n_{0j}\,,
\end{equation}
with $\gamma_j$ arbitrary real constants. To summarize, we found that the marginals are equal to: 
\begin{equation}
P_{j}(m_{j},\boldsymbol{n}_{0})=\begin{cases}
\delta_{m_{j},\gamma_{j}}\\
\delta_{m_{j},\gamma_{j}-2n_{0j}}\,,
\end{cases}\label{Pj_final}
\end{equation}
and since they are Kronecker deltas, the joint distribution is simply
\begin{equation}
P(\boldsymbol{m},\boldsymbol{n}_{0})=\prod_{j=1}^{3}P_{j}(m_{j},\boldsymbol{n}_{0})\label{Pfinal}\,.
\end{equation}

In order to establish how this requirement constrains the Kraus map in Eq. (1) in the main text, we recall that:
\begin{equation}
P(\boldsymbol{m},\boldsymbol{n})=\sum_{k}|\langle\boldsymbol{m}+\boldsymbol{n}|\hat{A}_{k}|\boldsymbol{n}\rangle|^{2}
\end{equation}
and from all analysis above
\begin{equation}
P(\boldsymbol{m},\boldsymbol{n})=\delta_{\boldsymbol{m},\boldsymbol{\gamma}(\boldsymbol{n})}\;\;\;\textrm{where}\;\;\;\gamma_{j}(n_{j})=\begin{cases}
\gamma_{j}\\
\gamma_{j}-2n_{j}
\end{cases},\label{alpha}
\end{equation}
which imply that
for each $k$
\begin{equation}
|\langle\boldsymbol{m}+\boldsymbol{n}|\hat{A}_{k}|\boldsymbol{n}\rangle|^{2}=c_{k}(\boldsymbol{n})\delta_{\boldsymbol{m},\boldsymbol{\gamma}(\boldsymbol{n})}\,,\label{condimento}
\end{equation}
with $\sum_{k}c_{k}(\boldsymbol{n})=1$, $c_{k}(\boldsymbol{n})\geq0$.
By decomposing the matrix element as
\begin{equation}
\langle\boldsymbol{m}+\boldsymbol{n}|\hat{A}_{k}|\boldsymbol{n}\rangle=R_{k}(\boldsymbol{m},\boldsymbol{n})e^{i\varphi_{k}(\boldsymbol{m},\boldsymbol{n})}\,,
\end{equation}
one finds that the condition~\eqref{condimento} does not restrict $\varphi_{k}(\boldsymbol{m},\boldsymbol{n})$
but implies
\begin{equation}
R_{k}(\boldsymbol{m},\boldsymbol{n})=\sqrt{c_{k}(\boldsymbol{n})}\delta_{\boldsymbol{m},\boldsymbol{\gamma}(\boldsymbol{n})}.
\end{equation}
Accordingly, the Kraus operator can be expressed as:
\begin{equation}
\hat{A}_{k}=\sum_{\boldsymbol{m},\boldsymbol{n}}|\boldsymbol{m}+\boldsymbol{n}\rangle\langle\boldsymbol{m}+\boldsymbol{n}|A_{k}|\boldsymbol{n}\rangle\langle\boldsymbol{n}|=\sum_{\boldsymbol{n}}\sqrt{c_{k}(\boldsymbol{n})}|\boldsymbol{\gamma}(\boldsymbol{n})+\boldsymbol{n}\rangle\langle\boldsymbol{n}|e^{i\varphi_{k}(\boldsymbol{\gamma}(\boldsymbol{n}),\boldsymbol{n})}\,,\label{Kraus op final}
\end{equation}
which once replaced in Eq. (4) in the main text gives
\begin{equation}
\hat{A}_{k}(\boldsymbol{x})=\sum_{\boldsymbol{n}}\sqrt{c_{k}(\boldsymbol{n})}|\boldsymbol{\gamma}(\boldsymbol{n})+\boldsymbol{n}\rangle\langle\boldsymbol{n}|e^{i\varphi_{k}(\boldsymbol{\gamma}(\boldsymbol{n}),\boldsymbol{n})}e^{\frac{2\pi i}{L}\boldsymbol{\gamma}(\boldsymbol{n})\cdot\boldsymbol{x}}
\end{equation}
and the translation covariant CP map reads 
\[
\varPhi[\hat{\rho}]=\frac{1}{L^{3}}\int_{-\frac{L}{2}}^{+\frac{L}{2}}d\boldsymbol{x}\sum_{k}\hat{A}_{k}(\boldsymbol{x})\hat{\rho} \hat{A}_{k}^{\dagger}(\boldsymbol{x})=
\]
\[
=\!\sum_{\boldsymbol{n},\boldsymbol{\ell}}\!\sum_{k}\sqrt{c_{k}(\boldsymbol{n})c_{k}(\boldsymbol{\ell})}e^{i[\varphi_{k}(\boldsymbol{\gamma}(\boldsymbol{n}),\boldsymbol{n})-\varphi_{k}(\boldsymbol{\gamma}(\boldsymbol{\ell}),\boldsymbol{\ell})]}\left(\frac{1}{L^{3}}\!\int_{-\frac{L}{2}}^{+\frac{L}{2}}\!\!d\boldsymbol{x}\!e^{\frac{2\pi i}{L}[\boldsymbol{\gamma}(\boldsymbol{n})-\boldsymbol{\gamma}(\boldsymbol{\ell})]\cdot\boldsymbol{x}}\!\right)\!\!\langle\boldsymbol{n}|\hat{\rho}|\boldsymbol{\ell}\rangle|\boldsymbol{\gamma}(\boldsymbol{n})+\boldsymbol{n}\rangle\langle\boldsymbol{\gamma}(\boldsymbol{\ell})+\boldsymbol{\ell}|.
\]
\begin{equation}
=\sum_{\boldsymbol{n},\boldsymbol{\ell}}\sum_{k}\sqrt{c_{k}(\boldsymbol{n})c_{k}(\boldsymbol{\ell})}e^{i[\varphi_{k}(\boldsymbol{\gamma}(\boldsymbol{n}),\boldsymbol{n})-\varphi_{k}(\boldsymbol{\gamma}(\boldsymbol{\ell}),\boldsymbol{\ell})]}\left(\delta_{\boldsymbol{\gamma}(\boldsymbol{n}),\boldsymbol{\gamma}(\boldsymbol{\ell})}\right)\langle\boldsymbol{n}|\hat{\rho}|\boldsymbol{\ell}\rangle|\boldsymbol{\gamma}(\boldsymbol{n})+\boldsymbol{n}\rangle\langle\boldsymbol{\gamma}(\boldsymbol{\ell})+\boldsymbol{\ell}|.
\end{equation}
Since according to Eq. (\ref{alpha}) each $\gamma_{j}(n_{j})$ can take two
possible values we have several possibilities. However, this is not
really important for us since, if we consider as initial state a plane
wave $\hat{\rho}=|\boldsymbol{n}_{0}\rangle\langle\boldsymbol{n}_{0}|$,
which is the most delocalized state in space, we have: 
\begin{equation}
\varPhi[\hat{\rho}]=|\boldsymbol{\gamma}(\boldsymbol{n}_{0})+\boldsymbol{n}_{0}\rangle\langle\boldsymbol{\gamma}(\boldsymbol{n}_{0})+\boldsymbol{n}_{0}|\label{super final}
\end{equation}
which means the dynamics does not collapse plane waves, it just give a boost $\boldsymbol{\gamma}(\boldsymbol{n}_{0})$. This conclude our proof. 

 A final comment about the relation between this theorem and Heisenberg's uncertainty principle might be useful.  The spread can be computed either at the wave function level, or at the density matrix level; for a stochastic dynamics---as typical for collapse models---the first case refers to a single realization of the noise, while  the second case refers to the average over all realizations, and is the one which is associated to experiments. Here we are interested in this second case.
 
 At the density matrix level, the spread in position in general does not decrease after the collapse (whose main effect is to cancel the off-diagonal elements of the density matrix in the position basis, not the diagonal ones), actually it increases more than what expected  by the Schr\"odinger's dynamics alone \cite{bassi2003dynamical}. As such, Heisenberg's principle does not require  the spread in momentum to increase after the collapse; yet our theorem shows that it must, under the specified assumptions.
 
 Even at the wave function level, it is not true that a collapse in position increases the spread in momentum, unless the wave function starts in a state of minimum uncertainty. What happens in general \cite{bassi2005collapse, bassi2005energy} is that a collapse in position also localizes the wave function in momentum, so that any initial state converges asymptotically to a state with (almost) the minimum uncertainty allowed by Heisenberg's uncertainty principle.

\subsection{Proof of the theorem for a Lindblad dynamics}\label{Sec_Lind}

Here we prove the theorem for the case of CP and space-translation covariant Quantum Dynamical Semigroup $\{\varPhi_t,t\geq0\}$, whose generator is of the Lindblad type~\cite{gorini1976completely, lindblad1976generators}. According to Holevo's theorem \cite{holevo1993note,holevo1993conservativity,vacchini2008kinetic,vacchini2007precise,vacchini2009quantum}, it takes the form  $\dot{\hat{\rho}}(t)=-\frac{i}{\hbar}[\hat H,\hat\rho(t)]+\mathcal{L}[\hat{\rho}(t)]$, with 
\begin{equation}
\mathcal{L}[\hat{\rho}(t)] = \int d\boldsymbol{Q}\sum_{j=1}^{\infty}\left(e^{\frac{i}{\hbar}\boldsymbol{Q}\cdot\hat{\boldsymbol{x}}} \hat L_{j}\hat{\rho}(t) \hat L_{j}^{\dagger}e^{-\frac{i}{\hbar}\boldsymbol{Q}\cdot\hat{\boldsymbol{x}}}-\frac{1}{2}\left\{ \hat L_{j}^{\dagger} \hat L_{j},\hat{\rho}(t)\right\} \right),\label{holev_lind}
\end{equation}
where we used the shorthand notation $\hat L_{j} = \hat L_{j}(\boldsymbol{Q},\hat{\boldsymbol{p}})$, and such operators satisfy
\begin{equation}\label{nodiv}
\int d\boldsymbol{Q}\sum_{j=1}^{\infty}|\hat L_{j}(\boldsymbol{Q},\cdot)|^{2}<\infty.
\end{equation}
We recall that we are considering a single particle or, alternatively, the center of mass of a composite object. The Hamiltonian evolution might change the spread of the particle in momentum, but here we are interested only in the diffusive contribution given by $\mathcal{L}[\hat{\rho}(t)]$. 

Similarly to what we did above, we carry out the calculation by confining the system in a box of size $L$ with periodic boundary conditions. This implies that in Eqs. (\ref{holev_lind}) and (\ref{nodiv}) the variable $\boldsymbol{Q}$ is discrete: $\boldsymbol{Q}=(2\pi\hbar/L) \boldsymbol{\ell}:=\tilde{\boldsymbol{\ell}}$;  integration over $\boldsymbol{Q}$ is replaced by a sum over $\boldsymbol{\ell}\in\mathbb{Z}^{3}$ and, as before,  the eigenvalues of the momentum operator take only discrete values $\boldsymbol{p}=(2\pi\hbar/L)\boldsymbol{n}:=\tilde{\boldsymbol{n}}$ with $\boldsymbol{n}\in\mathbb{Z}^{3}$. 

Then Eq. (\ref{holev_lind}) becomes:
\begin{equation}\label{holevo_lindbo}
    \mathcal{L}[\hat{\rho}(t)]=\left(\frac{2\pi\hbar}{L}\right)^{3}\sum_{\boldsymbol{\ell}}\sum_{j=1}^{\infty}\left(e^{\frac{i}{\hbar}\tilde{\boldsymbol{\ell}}\cdot\hat{\boldsymbol{x}}}\hat{L}_{j}\hat{\rho}(t)\hat{L}_{j}^{\dagger}e^{-\frac{i}{\hbar}\tilde{\boldsymbol{\ell}}\cdot\hat{\boldsymbol{x}}}-\frac{1}{2}\left\{ \hat{L}_{j}^{\dagger}\hat{L}_{j},\hat{\rho}(t)\right\} \right)
\end{equation}
with $\hat{L}_{j}=\hat{L}_{j}(\tilde{\boldsymbol{\ell}},\hat{\boldsymbol{p}})$.

A straightforward calculation, making use of the cyclicity of the trace and of the identity $e^{-\frac{i}{\hbar}\boldsymbol{Q}\cdot\hat{\boldsymbol{x}}}\hat{\boldsymbol{p}}e^{\frac{i}{\hbar}\boldsymbol{Q}\cdot\hat{\boldsymbol{x}}}=\hat{\boldsymbol{p}}+\boldsymbol{Q}$,
leads to
\begin{equation}
\textrm{Tr}\left[\hat{\boldsymbol{p}}\mathcal{L}[\hat{\rho}(t)]\right]\!=\left(\frac{2\pi\hbar}{L}\right)^{3}\sum_{\boldsymbol{\ell}}\sum_{\boldsymbol{n}}f(\tilde{\boldsymbol{\ell}},\tilde{\boldsymbol{n}})\tilde{\boldsymbol{\ell}}\langle\boldsymbol{n}|\hat{\rho}(t)|\boldsymbol{n}\rangle\label{p_lind-2}
\end{equation}
and 
\begin{equation}
\textrm{Tr}\left[\hat{\boldsymbol{p}}^{2}\!\mathcal{L}[\hat{\rho}(t)]\right]\!=\!\left(\frac{2\pi\hbar}{L}\right)^{3}\sum_{\boldsymbol{\ell}}\sum_{\boldsymbol{n}}f(\tilde{\boldsymbol{\ell}},\tilde{\boldsymbol{n}})(\tilde{\boldsymbol{\ell}}^{2}+2\tilde{\boldsymbol{n}}\cdot\tilde{\boldsymbol{\ell}})\langle\boldsymbol{n}|\hat{\rho}(t)|\boldsymbol{n}\rangle,\label{p2_lind-2}
\end{equation}
where we defined 
\begin{equation}
f(\tilde{\boldsymbol{\ell}},\tilde{\boldsymbol{n}}):=\sum_{j=1}^{\infty}|L_{j}(\tilde{\boldsymbol{\ell}},\tilde{\boldsymbol{n}})|^{2}\label{f_lind-1},
\end{equation}
with $L_{j}(\tilde{\boldsymbol{\ell}},\tilde{\boldsymbol{n}})$ eigenvalues of the $\hat{L}_j$ operators i.e. $\hat{L}_{j}(\tilde{\boldsymbol{\ell}},\hat{\boldsymbol{p}})|\boldsymbol{n}\rangle=L_{j}(\tilde{\boldsymbol{\ell}},\tilde{\boldsymbol{n}})|\boldsymbol{n}\rangle$.

Similarly to the proof in the main text, we assume that $\textrm{Tr}\left[\hat{\boldsymbol{p}}\mathcal{L}[\hat{\rho}(t)]\right]\!=0$
for any $\hat{\rho}(t)$. Accordingly, Eq. (\ref{p_lind-2}) implies 
$\sum_{\boldsymbol{\ell}}f(\tilde{\boldsymbol{\ell}},\tilde{\boldsymbol{n}})\tilde{\boldsymbol{\ell}}=0$ and Eq. (\ref{p2_lind-2}) reduces to 
\begin{equation}
\textrm{Tr}\left[\hat{\boldsymbol{p}}^{2}\mathcal{L}[\hat{\rho}(t)]\right]\!=    \left(\frac{2\pi\hbar}{L}\right)^{3}\sum_{\boldsymbol{\ell}}\sum_{\boldsymbol{n}}f(\tilde{\boldsymbol{\ell}},\tilde{\boldsymbol{n}})\tilde{\boldsymbol{\ell}}^{2}\langle\boldsymbol{n}|\hat{\rho}(t)|\boldsymbol{n}\rangle.
\end{equation}
The condition of having no diffusion, i.e. $\textrm{Tr}\left[\hat{\boldsymbol{p}}^{2}\mathcal{L}[\hat{\rho}(t)]\right]=0$
for any $\hat{\rho}(t)$, implies 
\begin{equation}
\sum_{\boldsymbol{\ell}}f(\tilde{\boldsymbol{\ell}},\tilde{\boldsymbol{n}})\tilde{\boldsymbol{\ell}}^{2}=0.\label{final_lind-1}
\end{equation}
Since $f(\tilde{\boldsymbol{\ell}},\tilde{\boldsymbol{n}})$ is, by definition, a positive function, Eq.~\eqref{final_lind-1} implies $f(\tilde{\boldsymbol{\ell}},\tilde{\boldsymbol{n}})=\lambda(\tilde{\boldsymbol{n}}) (2\pi\hbar/L)^{-3}\delta_{\boldsymbol{\ell},0}$ for all $\tilde{\boldsymbol{n}}$ (the factor $(2\pi\hbar/L)^{-3}$ is necessary to obtain a well defined limit when $L\rightarrow\infty$ in the final result in Eq. (\ref{ffff})). 
Given Eq. (\ref{f_lind-1}), this implies 
\begin{equation}
    L_{j}(\tilde{\boldsymbol{\ell}},\tilde{\boldsymbol{n}})=|L_{j}(\tilde{\boldsymbol{\ell}},\tilde{\boldsymbol{n}})|e^{i\varphi_{j}(\tilde{\boldsymbol{\ell}},\tilde{\boldsymbol{n}})}=\sqrt{\lambda_{j}(\tilde{\boldsymbol{n}})}\left(\frac{2\pi\hbar}{L}\right)^{-\frac{3}{2}}\delta_{\boldsymbol{\ell},0}e^{i\varphi_{j}(\tilde{\boldsymbol{n}})}
\end{equation}
with $\sum_j \lambda_j(\tilde{\boldsymbol{n}})=\lambda(\tilde{\boldsymbol{n}})$. From this it follows that the Lindbladian in Eq. (\ref{holevo_lindbo}) is of the form:
\begin{equation}\label{ffff}
   \mathcal{L}[\hat{\rho}(t)]=\sum_{j=1}^{\infty}\left(\hat{L}_{j}(\hat{\boldsymbol{p}})\hat{\rho}(t)\hat{L}_{j}^{\dagger}(\hat{\boldsymbol{p}})-\frac{1}{2}\left\{ \hat{L}_{j}^{\dagger}(\hat{\boldsymbol{p}})\hat{L}_{j}(\hat{\boldsymbol{p}}),\hat{\rho}(t)\right\} \right),
\end{equation}
with $\hat{L}_{j}(\hat{\boldsymbol{p}})=\sum_{\boldsymbol{n}}\sqrt{\lambda_{j}(\tilde{\boldsymbol{n}})}e^{i\varphi_{j}(\tilde{\boldsymbol{n}})}|\boldsymbol{n}\rangle\langle\boldsymbol{n}|$.

This result is consistent with what found in Eq. (17) in the main text. 

\end{widetext}

\end{document}